\documentclass[preprint]{aastex}
\usepackage{psfig}
\slugcomment{KSUPT-02/5 \hspace{0.5truecm} September 2002}

\newcommand{\wisk}[1]{\ifmmode{#1}\else{$#1$}\fi}

\def\lsim{\;_\sim^<\;}

\begin{document}

\title{CMB Anisotropy Constraints on Flat-$\Lambda$ and Open CDM Cosmogonies 
       from DMR, UCSB South Pole, Python, ARGO, MAX, White Dish, OVRO, and 
       SuZIE Data}

\author{
  Pia~Mukherjee\altaffilmark{1},
  Ken~Ganga\altaffilmark{2},
  Bharat~Ratra\altaffilmark{1}, 
  Gra{\c c}a~Rocha\altaffilmark{1,3},
  Tarun~Souradeep\altaffilmark{1,4},
  Naoshi~Sugiyama\altaffilmark{5},
  and
  Krzysztof~M.~G\'orski\altaffilmark{6}
  }

\altaffiltext{1}{Department of Physics, Kansas State University,
                 116 Cardwell Hall, Manhattan, KS 66506.}
\altaffiltext{2}{IPAC, MS 100--22, California Institute of Technology,
                 Pasadena, CA 91125.}
\altaffiltext{3}{Current address: Cavendish Laboratory,
                 University of Cambridge, Madingley Road, Cambridge, 
                 CB3 0HE, UK.}
\altaffiltext{4}{Current address: IUCAA, Post Bag 4, Ganeshkhind, Pune 
                 411007, India.}
\altaffiltext{5}{Division of Theoretical Astrophysics, National Astronomical
                 Observatory, 2-21-1 Osawa, Mitaka, Tokyo 181-8588, Japan;
                 and Max Planck Institute for Astrophysics, Karl-Schwarzschild 
                 Str. 1, Postfach 1317, Garching D-85741, Germany.}
\altaffiltext{6}{European Southern Observatory, Karl-Schwarzschild Str. 2, 
                 Garching D-85748, Germany; and Warsaw University Observatory, 
                 Aleje Ujazdowskie 4, 00-478 Warszawa, Poland.}

\begin{abstract}
We use joint likelihood analyses of combinations of fifteen cosmic microwave 
background (CMB) anisotropy data sets from the DMR, UCSB South Pole 1994, 
Python I--III, ARGO, MAX 4 and 5, White Dish, OVRO, and SuZIE experiments 
to constrain cosmogonies. We consider open and spatially-flat-$\Lambda$ cold 
dark matter cosmogonies, with nonrelativistic-mass density parameter 
$\Omega_0$ in the range 0.1--1, baryonic-mass density parameter $\Omega_B$ 
in the range (0.005--0.029)$h^{-2}$, and age of the universe $t_0$ in the 
range (10--20) Gyr. 

Marginalizing over all parameters but $\Omega_0$, the data favor
$\Omega_0\simeq$ 0.9--1 (0.4--0.6) flat-$\Lambda$ (open) models. The 
range in deduced $\Omega_0$ values is partially a consequence of the 
different combinations of smaller-angular-scale CMB anisotropy data sets 
used in the analyses, but more significantly a consequence of whether
the DMR quadrupole moment is accounted for or ignored in the analysis. 
While the open model is difficult to reconcile with the results of less 
exact analyses of more recent CMB anisotropy data, the lower values of 
$\Omega_0$ found in this case are more easily reconciled with dynamical
estimates of this parameter.
For both flat-$\Lambda$ and open models, after marginalizing over all other 
parameters, a lower $\Omega_B h^2 \simeq$ 0.005--0.009 is favored. This
is also marginally at odds with estimates from more recent CMB anisotropy 
data and some estimates from standard nucleosynthesis theory and observed 
light element abundances. For both sets of models a younger universe 
with $t_0 \simeq$ 12--15 Gyr is favored, consistent with other recent 
non-CMB indicators. We emphasize that since we consider only a small 
number of data sets, these results are tentative. More importantly, the
analyses here do not rule out the currently favored flat-$\Lambda$ model
with $\Omega_0 \sim 0.3$, nor the larger $\Omega_B h^2$ values favored
by some other data.
\end{abstract}


\section{Introduction}

There has been a remarkable increase in the quality and quantity of 
cosmic microwave background (CMB) anisotropy measurements since the
initial detection of the anisotropy on large angular scales a decade 
ago.\footnote{
See, e.g., Miller et al. (2002a), Coble et al. (2001), Scott et al. (2002),
and Mason et al. (2002) for recent measurements.}
These measurements are becoming increasingly useful in the continuing
processes of determining how well cosmological models approximate reality
and for constraining cosmological parameters such as $\Omega_0$, $h$, and 
$\Omega_B$ in these models\footnote{
Here $h$ is the Hubble constant $H_0$ in units of $100\ {\rm km}\ 
{\rm s}^{-1}\ {\rm Mpc}^{-1}$. See, e.g., Podariu et al. (2001), 
Wang, Tegmark, \& Zaldarriaga (2002), Miller et al. (2002b), Sievers et al.
(2002), and Lewis \& Bridle (2002) for recent discussions of constraints on 
cosmological parameters.}.

In this paper we utilize the full information in the data from each 
experiment in an effort to place robust constraints on cosmological 
parameters. This is achieved through a maximum likelihood analysis of 
the data using realistic model anisotropy spectra. Most of the data 
sets we consider in this paper are small, in contrast to some of 
the more recent data sets (see, e.g., Netterfield et al. 2002; Stompor 
et al. 2001), which because of their size require a more approximate 
analysis technique.
Ganga et al. (1997a, hereafter GRGS) extend the maximum likelihood
technique to account for uncertainties, such as those in the beamwidth 
of the telescope and the calibration of the experiment.\footnote{
For some of the data sets we consider in this paper foreground non-CMB
contamination must also be accounted for; see, e.g., Kogut et al. (1996),
de Oliveira-Costa et al. (1998), Hamilton \& Ganga (2001), and Mukherjee 
et al. (2002a, 2002b) for discussions of the method used to accomplish this.}
This technique has been used with model CMB anisotropy spectra in analyses 
of the Gundersen et al. (1995) UCSB South Pole 1994 data, the Church et 
al. (1997) SuZIE data, the Lim et al. (1996) MAX 4+5 data, the Tucker et 
al. (1993) White Dish data, the de Bernardis et al. (1994) ARGO data, the 
Platt et al. (1997) Python I--III data, and the Leitch et al. (2000) OVRO 
data (GRGS; Ganga et al. 1997b, 1998; Ratra et al. 1998, 1999a, hereafter 
R99a; Rocha et al. 1999; Mukherjee et al. 2002b).

Given the error bars associated with these measurements, interesting 
constraints on cosmological model parameters require the joint analysis
of many data sets. If the measurements are acquired for regions well
separated in space, or on very different angular scales, the likelihoods 
of the individual data sets are independent and can thus be multiplied 
together to construct the likelihood of the combined data set. This 
combined likelihood is then used to derive constraints on cosmological 
model parameters.  A combined analysis of the smaller-angular-scale data 
sets listed above, excluding the Python and OVRO data, is presented 
in Ratra et al. (1999b, hereafter R99b).
In this paper we extend the analysis of R99b to include the Python and OVRO 
data, as well as the large angular scale DMR data (G\'orski et al. 1998,
hereafter G98; Stompor 1997).

In $\S$ 2 we describe the models and cosmological parameter space we 
consider. See R99a for a more detailed description. In $\S$ 3 we summarize
the various combinations of data sets we consider. See R99b for further
details. In $\S$ 4 we summarize the computational techniques we use. See 
GRGS and R99b for more detailed discussions. In $\S$ 5 we present and 
discuss results from analyses of various combinations of the 
smaller-angular-scale CMB anisotropy data sets, and in $\S$ 6 we add 
the DMR data to the mix. We conclude in $\S$ 7.

\section{Cosmogonical Models}

In this paper we focus on a spatially-flat cold dark matter (CDM) model 
with a cosmological constant $\Lambda$. This model is consistent
with most current measurements.\footnote{
See, e.g., Peebles (1984), Efstathiou, Sutherland, \& Maddox (1990), 
Stompor, G\'orski, \& Banday (1995), Ratra et al. (1997), Sahni \& 
Starobinsky (2000), Carroll (2001), and Peebles \& Ratra (2002) for 
discussions of this model. While not considered in this paper, a 
spatially-flat model dominated at the current epoch by time-variable 
dark energy is also largely consistent with current measurements
(see, e.g., Peebles \& Ratra 1988; Ratra \& Quillen 1992; Steinhardt 1999;
Brax, Martin, \& Riazuelo 2000; Munshi \& Wang 2002; Bartelmann, Perrotta,
\& Baccigalupi 2002; Dave, Caldwell, \& Steinhardt 2002; Chen \& Ratra 2002;
Podariu et al. 2002).}
While it is important to compare the currently favored model to the CMB 
anisotropy data, it is also important to check how well other, currently 
less favored, models fare. We therefore also examine the constraints these
data place on a spatially-open model with no $\Lambda$ (see, e.g., Gott 
1982; Ratra \& Peebles 1995). For more detailed discussions of these models 
see R99a and R99b.

The CMB anisotropy spectra in these models are generated as quantum 
fluctuations in a weakly coupled field during an early epoch of inflation.
They are thus realizations of spatially stationary Gaussian random
processes (see, e.g., Ratra 1985; Fischler, Ratra, \& Susskind
1985). The measured CMB anisotropy appears to be consistent with this
Gaussianity assumption (see, e.g., Mukherjee, Hobson, \& Lasenby 2000;
Park et al. 2001; Wu et al. 2001; Shandarin et al. 2002; Santos et al. 
2002; Komatsu et al. 2002;
Polenta et al. 2002), and the experimental noise also appears to be Gaussian, thus validating our use of the likelihood analysis method of GRGS. 

To make the computations tractable, in each model (flat-$\Lambda$ and open)
we consider CMB anisotropy spectra parameterized by (i) the quadrupole-moment
amplitude $Q_{\rm rms-PS}$, (ii) $\Omega_0$, (iii) $\Omega_B h^2$, and 
(iv) $t_0$. While it is of interest to also consider 
other cosmological parameters current data do not require consideration
of a larger dimensional parameter space. That is, current data appear to 
indicate that effects like reionization and tilt and those due to primordial 
gravity waves are small. Furthermore, the maximum likelihood analysis 
technique used here is computationally intensive and an analysis that takes
account of more cosmological parameters will be very time consuming. Rather
than make approximations (e.g., data compression) to speed up such a 
computation and minimize memory requirements, in this paper we perform 
as accurate a computation as possible based on model spectra from an as small
as viable yet still realistic cosmological parameter space.

The CMB anisotropy spectra we use are computed for a range of 
$\Omega_0$ spanning the interval 0.1 to 1 in steps of 0.1, for a range of 
$\Omega_B h^2$ spanning the interval 0.005 to 0.029 in steps of 0.004, 
and for a range of $t_0$ spanning the interval 10 to 20 Gyr in steps of 2 
Gyr. In total 798 spectra are computed to cover the cosmological-parameter 
spaces of the open and flat-$\Lambda$ models. Examples of spectra are shown 
in Fig.~2 of R99a, Fig.~1 of R99b, and Fig.~2 of Rocha et al.~(1999).

\section{CMB Anisotropy Data Sets}

R99b consider various combinations of ten different data sets. 
These were the UCSB South Pole 1994 Ka and Q band observations,
hereafter SP94Ka and SP94Q (Gundersen et al. 1995; GRGS), the ARGO Hercules
observations (de Bernardis et al. 1994; R99a), the MAX 4 $\iota$ Draconis
(ID) and $\sigma$ Herculis (SH) and MAX 5 HR5127 (HR), $\mu$ Pegasi (MP),
and $\phi$ Herculis (PH) observations (Lim et al. 1996; Ganga et al. 1998), 
the White Dish observations (Tucker et al. 1993; Ratra et al. 1998), and the 
SuZIE observations (Church et al. 1997; Ganga et al. 1997b). Detailed 
information about these data sets may be found in the papers cited above
and in R99b.

In this paper we augment the data sets used in R99b with the Python I--III
data (Platt et al. 1997; Rocha et al. 1999), the OVRO data (Leitch et al. 
2000; Mukherjee et al. 2002b), and the DMR data (Bennett et al. 1996; G98; 
Stompor 1997). This adds an additional five CMB anisotropy data sets to the 
ten considered in R99b.

R99b consider three different combinations of the CMB anisotropy
data sets. The first combination included all data (SP94, ARGO, MAX 4 and 5, 
White Dish, and SuZIE). The second combination excluded the SuZIE data, 
which probe the smallest angular scales (to multipole index $l \sim 2000$
in some models, Ganga et al. 1997b), since our model CMB anisotropy spectra 
do not account for effects that could be important on these small
angular scales. The third combination of data used in R99b include just
the measurements of SP94Ka, MAX 4 ID, and MAX 5 HR which are thought to
provide the most reliable constraints, among the data considered in R99b
(see discussion in R99b), to constrain the parameters of the theoretical 
CMB anisotropy spectra considered in that paper.

The new smaller-angular-scale data used in this paper, Python I--III and
OVRO, raise no obvious red flags\footnote{
We note that Mukherjee et al. (2002b) account for the removal of non-CMB
foreground contamination in the OVRO analysis.}: 
we have no reason to think that they 
are any less reliable than the data considered to be the most reliable 
in R99b. We therefore again derive cosmological constraints 
for three different combination CMB anisotropy data sets, augmenting each 
of the three combination data sets of R99b with the Python I--III and 
OVRO data.

The situation with the largest angular scale CMB anisotropy data, that
measured by the DMR experiment (Bennett et al. 1996; G98), is more 
complicated. While a couple of other issues also need to be considered,
the main complication arises from the effect on the derived cosmological
constraints from the inclusion or exclusion of the observed DMR quadrupole
moment (Kogut et al. 1996) in the analysis (G98; Stompor 1997). Non-CMB
anisotropy emission from the Milky Way is predominantly quadrupolar,
and any post-correction remnant of this will primarily affect the observed DMR
quadrupole moment. To be sure, this is only of order a 1 $\sigma$ effect
on cosmological parameter determination from the DMR data alone. However,
given the differences in the constraints derived from the different
combinations of smaller-angular-scale data sets we consider (and again, 
these differences are of order 1 $\sigma$), folding in the DMR data results
in more significant differences between cosmological constraints derived 
using the various combination data sets. One might hope that
better maps of foreground emission will help resolve the DMR quadrupole
moment issue, but that is for the future. The only option at present
appears to be to consider cosmological constraints that follow on using
all ``reasonable" combinations of data. To illustrate the situation we 
consider the two most extreme DMR data combinations (G98): (i) the galactic
frame data including the quadrupole moment and correcting for faint
high-latitude galactic emission (this results in the smallest
$Q_{\rm rms-PS}$ value); and (ii) the ecliptic frame data with
the quadrupole moment excluded and no other galactic emission correction
(this results in the largest $Q_{\rm rms-PS}$ value).
In conjunction with the three different combinations of smaller-angular-scale 
data, these two DMR data sets result in six different combinations of
CMB anisotropy data we consider in this paper.

\section{Summary of Computation}

GRGS describe the computation of the likelihood function for a 
given CMB anisotropy data set. Offsets and gradients removed from 
the data are accounted for in the analysis. Beamwidth and calibration
uncertainties are also accounted for as described in GRGS. In some cases
non-CMB anisotropy foreground contamination have also been accounted
for.

The data sets considered in this analysis were acquired from regions
well separated in space, or are at very different angular resolution.
Consequently they are statistically independent and the likelihood
functions of the individual data sets are simply multiplied together to 
construct the likelihood function of the combined data. The flat-$\Lambda$ 
and open model likelihoods we consider are a function of the four 
parameters described in $\S$ 2: $Q_{\rm rms-PS}$, $\Omega_0$, 
$\Omega_B h^2$, and $t_0$. Marginalized likelihood functions are derived by 
integrating over one or more of these parameters. We assume a uniform prior 
in the parameters integrated over, set to zero outside the parameter range 
considered.

To determine central values and limits from the likelihood functions we 
assume a uniform prior in the relevant parameter. Then the corresponding 
posterior probability density distribution function vanishes outside the
chosen parameter range and is equal to the likelihood function inside
this range. The deduced central value of the parameter is taken to be the 
value at which the posterior probability density peaks, and we quote highest 
posterior density limits. See GRGS, R99a, and R99b for details. The 
quoted limits depend on the prior range considered for the parameter.
This is a significant effect if the likelihood function is not sharply peaked 
within the parameter range considered, as is the case for a number of the 
likelihood functions derived in this paper. See R99a and R99b for more 
detailed discussions of this issue. In what follows we consider 1, 2, and
3 $\sigma$ highest posterior density limits which include 68.3, 95.4, and
99.7\% of the probability.

Rocha et al. (1999) and Mukherjee et al. (2002b) note that the 
four-dimensional posterior probability density distribution function 
$L(Q_{\rm rms-PS}, \Omega_0, \Omega_B h^2, t_0)$ is nicely peaked in 
the $Q_{\rm rms-PS}$ direction but fairly flat in the other 
three directions. Marginalizing over $Q_{\rm rms-PS}$ results in a 
three-dimensional posterior distribution $L(\Omega_0,  \Omega_B h^2, t_0)$ 
which is steeper, but still relatively flat. As a consequence, limits 
derived from the four- and three-dimensional posterior distributions are
generally not highly statistically significant. We therefore do not show 
contour plots of these functions here. Marginalizing over $Q_{\rm rms-PS}$ and 
one other parameter results in two-dimensional posterior probability 
distributions which are more peaked. See Fig.~1 for examples of the 
constraints from the two-dimensional posterior probability distribution
functions. As in the ARGO (R99a), Python I--III (Rocha et al. 1999), 
OVRO (Mukherjee et al. 2002b), and earlier combination (R99b) data set 
analyses, in some cases these peaks are at an edge of the parameter range considered.

\section{Smaller Angular Scale Data: Results and Discussion}

Figure 1 shows that even just using the smaller angular scale CMB anisotropy
data, the two-dimensional posterior distributions allow one to 
distinguish between different regions of parameter space at a fairly high 
formal level of confidence.\footnote{
See Fig.~4 of R99a, Fig.~2 of R99b, Fig.~3 of Rocha et al. (1999), and
Fig.~1 of Mukherjee et al. (2002b) for related cosmological constraints 
from subsets of the data considered here.}
For instance, for the SP94Ka, Python I--III, MAX 4 ID, MAX 5 HR, and 
OVRO data combination (panels $c$ and $f$ in the bottom row of
Fig.~1), the flat-$\Lambda$ model near $\Omega_0 \sim 0.1$, $\Omega_B h^2 
\sim 0.03$, and $t_0 \sim 10$ Gyr, is formally ruled out at $\sim 3$ 
$\sigma$ confidence. However, we emphasize, as discussed in Mukherjee et al.
(2002b), care must be exercised when interpreting the discriminative power 
of these formal limits, since they depend sensitively on the fact that 
the uniform prior has been set to zero outside the range of the parameter 
space we have considered.

Figure 2 shows the contours of the two-dimensional posterior distribution 
for $Q_{\rm rms-PS}$ and $\Omega_0$, derived by marginalizing the 
four-dimensional distribution over $\Omega_B h^2$ and $t_0$. These are shown
for the different combinations of smaller-angular-scale data and for the DMR 
data, for both the open and flat-$\Lambda$ models. Constraints on cosmological
parameters from the smaller-angular-scale data are consistent with those 
from the DMR data.\footnote{
See Fig.~5 of R99a, Fig.~3 of R99b, Fig.~4 of Rocha et al. (1999) and 
Fig.~2 of Mukherjee et al. (2002b) for related constraints from subsets
of the data considered in this paper.}
It is interesting that the constraints on $Q_{\rm rms-PS}$ derived from the 
DMR data are not very much tighter than those derived from the different 
combinations of smaller-angular-scale data.

Figure 3 shows the one-dimensional marginalized posterior distribution 
functions for $\Omega_0$, $\Omega_B h^2$, and $t_0$. For the flat-$\Lambda$ 
model (first column of Fig.~3), all three combinations of data sets favor 
$\Omega_0 = 1$, and low values of $\Omega_0 \sim 0.1$ are disfavored, at 
least by those data set combinations that include most of the 
data. This should be compared to the earlier analysis of Fig.~5 of R99b 
(which did not include the Python I--III and OVRO data considered here),
where the data were not able to significantly discriminate between
different $\Omega_0$ values in the flat-$\Lambda$ case. For the open model 
(second column of Fig.~3) all three combinations of smaller-angular-scale
data sets favor $\Omega_0 \simeq 0.4$, with low values of $\Omega_0 \sim 0.1$ 
again disfavored. This should also be compared to the results shown in 
Fig.~5 of R99b, where it was found that conclusions about the favored value
of $\Omega_0$ in the open model were very dependent on the combination
of smaller-angular-scale data used to derive them. Apparently we now 
have a large enough collection of data sets for this to no longer be an issue.

The central two columns of Fig.~3 show the one-dimensional marginalized 
posterior distribution functions for $\Omega_B h^2$. Independent of 
the data set combination considered low values of $\Omega_B h^2$ are
favored, with $\Omega_B h^2 \simeq$ 0.005 (0.007--0.008) favored in the 
open (flat-$\Lambda$) model. These results are statistically much more 
significant than, but quite consistent with, those shown in Fig.~6 of R99b.

The last two columns of Fig.~3 show the one-dimensional marginalized posterior 
distribution functions for $t_0$. Independent of the data set combination considered the flat-$\Lambda$ case favors an old universe with $t_0 \simeq$
18--20 Gyr (fifth column of Fig.~3), while the open model favors a young 
universe with $t_0 \simeq$ 10--12 Gyr (sixth column of Fig.~3). These open 
model results are very consistent with those shown in Fig.~7 of R99b, while
R99b found that flat-$\Lambda$ models favored a young universe with 
$t_0 \simeq$ 10--12 Gyr, at odds with what we find here.

More precisely, the one-dimensional distributions for the ``most reliable"
data in Fig.~3 (the bottom row) indicates that an open (flat-$\Lambda$) 
model with $\Omega_0$ = 0.42 (1.0), or $\Omega_B h^2$ = 0.005 (0.008), 
or $t_0$ = 10 (18) Gyr is favored, among the models considered. At 
2 $\sigma$ confidence the data formally rule out only small regions of 
parameter space: the data require $\Omega_0$ $> 0.18$
($\Omega_0$ $> 0.10$), or $\Omega_B h^2$ $< 0.027$ 
($\Omega_B h^2$ $< 0.026$), or $t_0$ $< 19$ Gyr ($t_0$ $> 11$ Gyr) for the 
open (flat-$\Lambda$) model at 2 $\sigma$.

While the statistical significance of the constraints on cosmological 
parameters from these combinations of smaller-angular-scale CMB anisotropy
data is not high, there are a couple of puzzling issues. The lower 
$\Omega_0 \simeq 0.4$ favored in the open case is more easily reconciled 
with most other estimates of $\Omega_0$  than is the higher value of 
$\Omega_0 = 1$ favored by the CMB anisotropy data in the flat-$\Lambda$ model
(see, e.g., Peebles \& Ratra 2002). We emphasize however that more recent
CMB anisotropy data favor the flat-$\Lambda$ model with low $\Omega_0$ 
(see, e.g., Netterfield
et al. 2002; Pryke et al. 2002; Stompor et al. 2001; Scott et al. 2002;
Mason et al. 2002). The low values of $\Omega_B h^2$ favored by the data 
considered here are also somewhat at odds with the higher values favored 
by more 
recent CMB anisotropy data (see, e.g., Netterfield et al. 2002; Pryke et al. 
2002; Stompor et al. 2001) and those favored by standard nucleosynthesis
theory and the observed deuterium abundances (Burles, Nollett, \& Turner 
2001); they are, however, more consistent with the lower $\Omega_B h^2$ 
value from standard nucleosynthesis theory and the observed helium and 
lithium abundances (Cyburt, Fields, \& Olive 2001). The ages of an open 
and a flat-$\Lambda$ universe determined from the smaller-angular-scale 
CMB data used in this paper lie at the opposite ends of the 2 $\sigma$
range of estimates based on the ages of the oldest stars: 11 Gyr $\lsim
t_0 \lsim$ 17 Gyr (see, e.g., Carretta et al. 2000; Krauss \& Chaboyer 2001).
Given that the CMB anisotropy data considered here are not the most recent,
and that the statistical significance of the puzzling aspects of the 
derived results are not high, perhaps similar analyses of more recent and
future data will show that these puzzles are only apparent. 
 
Figure 4 shows the one-dimensional posterior distribution functions for
$Q_{\rm rms-PS}$, derived by marginalizing the four-dimensional 
ones over the other three parameters. The three different combinations
of smaller-angular-scale data sets considered here result in fairly 
tight constraints on $Q_{\rm rms-PS}$. At 2 $\sigma$ confidence they are consistent with the DMR results for both the open and flat-$\Lambda$ models.  

The peak values of the one-dimensional posterior distributions shown in 
Figs.~3 and 4 are listed in the figure captions for the case when the 
four-dimensional posterior distributions are normalized such that
$L(Q_{\rm rms-PS}\ =\ 0\ \mu{\rm K})\ =\ 1$. With this normalization, 
marginalizing over the remaining parameter the fully marginalized
posterior distributions are, for the open (flat-$\Lambda$) models:
$1\times 10^{272}(4\times 10^{271})$ for all the data;   
$1\times 10^{272}(2\times 10^{271})$ for all the data excluding SuZIE;
and, $1\times 10^{202}(8\times 10^{201})$ for the SP94Ka, Python I--III, 
MAX 4 ID, MAX 5 HR, and OVRO combination. These numerical values are 
consistent with the indications from the first two columns of Fig.~3
that the data mildly favor the open model over the flat-$\Lambda$ one
(since the $\Omega_0 = 1$ models at the right edges of pairs of panels 
in each row of Fig.~3 are identical).

\section{Including the DMR Data: Results and Discussion}

As mentioned above, cosmological parameter constraints derived from the 
DMR data are somewhat sensitive to whether the measured DMR quadrupole 
moment is included in or excluded from the analysis. To account for this 
source of uncertainty we consider the two most extreme (in terms of 
$Q_{\rm rms-PS}$ normalization) DMR data combinations (G98; Stompor 1997): 
(i) the galactic frame data including the quadrupole moment and correcting 
for faint high-latitude galactic emission; and (ii) the ecliptic frame data 
with the quadrupole moment excluded and no other galactic emission correction.
Since we consider three different combinations of smaller-angular-scale 
data, combining these two DMR data sets with the smaller-angular-scale
data results in the six different combinations of CMB anisotropy data we
consider in this section.

When the smaller-angular-scale data of $\S$ 5 are augmented with the 
DMR data, the four-dimensional posterior probability density distribution
function $L(Q_{\rm rms-PS}, \Omega_0, \Omega_B h^2, t_0)$ becomes 
much more peaked in the $t_0$ direction (in addition to being peaked in the 
$Q_{\rm rms-PS}$ direction) but is still somewhat flat in the other two 
directions.  As discussed in $\S$ 5, marginalizing over $Q_{\rm rms-PS}$ 
and one other parameter results in two-dimensional posterior probability 
functions which are more peaked. Examples are shown in Fig.~5. Comparing
to the corresponding plots (Fig.~1) based on just the smaller-angular-scale
data, one sees that the inclusion of the DMR data significantly tightens the 
constraints on cosmological parameters.  

For example, Fig.~5 shows that, independent of data set combination 
considered, flat-$\Lambda$ models with parameter values near $\Omega_0 
\sim 0.1$, $\Omega_B h^2 \sim 0.03$, and $t_0 \sim 20$ Gyr, and open 
models with $\Omega_0 \lsim 0.2$, are both formally ruled out at $\sim 3$ 
$\sigma$ confidence. 

Figure 6 shows the contours of the two-dimensional posterior distribution 
for $Q_{\rm rms-PS}$ and $\Omega_0$, derived by marginalizing the 
four-dimensional distribution over $\Omega_B h^2$ and $t_0$. These 
should be compared to the corresponding contours for just the 
smaller-angular-scale data which are shown in Fig.~2. The addition of the 
DMR data significantly tightens the constraints.

Figure 7 shows the one-dimensional posterior distribution functions for
$\Omega_0$, derived by marginalizing the four-dimensional ones over the 
other three parameters.\footnote{
We have checked in a couple of cases that the projected likelihood 
function is fairly similar to the marginalized one. This indicates that our
``prior" choice of setting to zero the likelihood function outside the 
parameter range we consider does not significantly affect our results.}
For the flat-$\Lambda$ model, using the galactic 
frame DMR data with quadrupole included and corrected for faint high-latitude
galactic emission, in conjunction with the smaller-angular-scale data,
results in $\Omega_0 \simeq 0.9$ being favored (first column of Fig.~7),
while the ecliptic frame DMR data with quadrupole excluded and no further
correction for foreground contamination results in $\Omega_0 = 1$ being
favored (third column of Fig.~7). In the open case, the galactic-frame
quadrupole-included DMR data, in conjunction with the smaller-angular-scale
data, favor $\Omega_0 \sim 0.5-0.6$ (second column of Fig.~7), while
the ecliptic-frame quadrupole-excluded DMR data set combination favor
$\Omega_0 \simeq 0.4$ (fourth column of Fig. 7). Perhaps the most reliable
constraints follow from panels $k$ and $l$ of Fig.~7, which are derived 
from the ecliptic frame DMR data with quadrupole excluded and no further
correction for foreground emission, used in conjunction with the 
smaller-angular-scale data from the SP94Ka, Python I--III, MAX 4 ID, MAX 5 HR,
and OVRO experiments. In this case, the favored value of $\Omega_0$ is
1.0 (0.43) with $\Omega > 0.17$ $(0.29 < \Omega_0 < 0.93)$ required at
2 $\sigma$ for the flat-$\Lambda$ (open) models.\footnote{
Clearly, the data considered here, which, aside from OVRO, are most sensitive
at multipole index $l < 200$ (OVRO is most sensitive at $l \sim 600$), 
favor a less steeply rising (with $l$) spectrum at $l < 200$, and also 
favor some significant amount of power at $l \sim 600$, thus favoring an open
model over $\Omega_0 = 1$ and the $\Omega_0 = 1$ model over a lower
density flat-$\Lambda$ one.}
Since the two $\Omega_0 = 1$ models at the right hand edges of each pair 
of plots are
identical, when the flat-$\Lambda$ and open model posterior probability
distribution functions are renormalized to make this $\Omega_0 = 1$  model
have the same probability in both cases, the probability for 
$\Omega_0 = 0.3$ (which is
close to what is observed, see, e.g., Peebles \& Ratra 2002) is about
50\% higher in the open case than in the flat-$\Lambda$ case. While
not highly statistically significant it is still interesting that the 
open ``foil" provides a better fit to this data than does the ``favored"
flat-$\Lambda$ model. It should also be noted that this conclusion also
follows from the smaller-angular-scale data alone: see the 
first two columns of Fig.~3. It is puzzling why these results are not
more consistent with those from less exact analysis of more recent 
CMB anisotropy data, which find that the low-density flat-$\Lambda$ model
is favored over the low-density open case (see, e.g., Netterfield et al.
2002; Pryke et al. 2002; Stompor et al. 2001). Perhaps the data used in
this paper has undetected systematic errors, or perhaps it is the more 
recent data which has these (we note that some of the more recent data 
analyses require use of ``prior" assumptions before they hone in on the 
``favored" flat-$\Lambda$ model). It appears that an analysis of more 
data than considered here will be needed to resolve this issue.

Figure 8 shows the one-dimensional marginalized posterior distribution 
functions for $\Omega_B h^2$. For both the flat-$\Lambda$ and open models, 
and for all combined data sets considered, a low $\Omega_B h^2 \simeq 
0.005-0.009$ is favored. For the most reliable data combination, the DMR 
ecliptic frame data with quadrupole excluded and no other correction for 
foreground emission, in conjunction with the SP94Ka, Python I-III, MAX 4 ID,
MAX 5 HR, and OVRO data sets (panels $k$ and $l$ of Fig. 8), $\Omega_B h^2
= 0.007$ is favored in both the flat-$\Lambda$ and open cases, with 
$\Omega_B h^2 < 0.026$ ($\Omega_B h^2 < 0.027$) required at 2 $\sigma$ for
the flat-$\Lambda$ (open) models. These values are somewhat lower than 
what are favored by less exact analyses of more recent CMB anisotropy 
data (see, e.g., Netterfield et al. 2002; Pryke et al. 2002; Stompor et al. 
2001) and those derived from standard nucleosynthesis theory and the 
observed deuterium abundances (Burles, Nollett, \& Turner 2001); they are, 
however, more consistent with the lower $\Omega_B h^2$ value from standard 
nucleosynthesis theory and the observed helium and lithium abundances 
(Cyburt, Fields, \& Olive 2001). These values are consistent with those 
derived using only the smaller-angular-scale data (see the central
two columns of Fig.~3).

Figure 9 shows the one-dimensional marginalized posterior distribution 
functions for $t_0$. The flat-$\Lambda$ model favors $t_0 \sim 12-15$ Gyr, 
while the open model favors $t_0 \sim 12-14$ Gyr. The open case is very 
consistent with the results from the smaller-angular-scale anisotropy
data alone (see the last column of Fig.~3), however, the smaller-angular-scale
data alone favors $t_0 \sim 18-20$ Gyr for the flat-$\Lambda$ model 
(see the second last column of Fig.~3), larger than what is favored when 
the DMR data is included in the mix. For the most reliable data combination, 
the DMR ecliptic frame data with quadrupole excluded and no other correction
for foreground emission, in conjunction with the SP94Ka, Python I-III,
MAX 4 ID, MAX 5 HR, and OVRO data (panels $k$ and $l$ of Fig.~9), $t_0
= 13$ Gyr is favored, with $t_0 < 19$ Gyr required at 2 $\sigma$ for 
both the open and flat-$\Lambda$ cases. These values are in excellent
accord with estimates for the age of the universe from the ages of the
oldest stars (see, e.g., Carretta et al. 2000; Krauss \& Chaboyer 2001).

Figure 10 shows the one-dimensional posterior distribution functions for
$Q_{\rm rms-PS}$, derived by marginalizing the four-dimensional 
ones over the other three parameters. The CMB anisotropy data lead
to tight constraints on $Q_{\rm rms-PS}$. 

The peak values of the one-dimensional posterior distributions shown in 
Figures 7--10 are listed in the figure captions for the case when the 
four-dimensional posterior distributions are normalized such that
$L(Q_{\rm rms-PS}\ =\ 0\ \mu{\rm K})\ =\ 1$. With this normalization, 
marginalizing over the remaining parameter the fully marginalized
posterior distributions are, for the flat-$\Lambda$ (open) models:
$3\times 10^{435}(7\times 10^{435})$ for the galactic-frame 
quadrupole-included faint-high-latitude-foreground-corrected DMR data
in conjunction with the smaller-angular-scale data; $1\times 10^{435}
(5\times 10^{435})$ for this data less the SuZIE data; 
$6\times 10^{365}(7\times 10^{365})$ for this DMR data and the SP94Ka, 
Python I-III, MAX 4 ID, MAX 5 HR, and OVRO smaller-angular-scale data;
$3\times 10^{440}(1\times 10^{441})$ for the ecliptic-frame 
quadrupole-excluded, and no other correction for foreground emission, 
DMR data in conjunction with the smaller-angular-scale data; $2\times 10^{440}
(1\times 10^{441})$ for this data less the SuZIE data; and, 
$1\times 10^{371}(2\times 10^{371})$ for this DMR data and the SP94Ka, 
Python I-III, MAX 4 ID, MAX 5 HR, and OVRO smaller-angular-scale data.
These numerical values are consistent with the indications from Fig.~7
that the open case is mildly favored over the flat-$\Lambda$ model.

\section{Conclusion}

We have derived constraints on cosmological model parameters in the open and 
flat-$\Lambda$ CDM models, from joint analyses of combinations of the DMR, 
SP94, Python I--III, ARGO, MAX 4+5, White Dish, OVRO, and SuZIE CMB 
anisotropy data sets. The constraints derived here are not of very high
statistical significance.

The data considered here mildly favor the open case over the flat-$\Lambda$
model, although they do not rule out the currently favored flat-$\Lambda$
model with $\Omega_0 \sim 0.3$. The favored value of $\Omega_0$ in the open 
($\Omega_0 \sim 0.4-0.6$) and flat-$\Lambda$ ($\Omega_0 \sim 0.9-1$) models 
are somewhat dependent on whether the DMR quadrupole moment is included in 
or excluded from the analysis. Resolving this issue will likely require 
analyses that use new higher quality large angular scale CMB anisotropy and 
foreground emission data. 

Constraints on $\Omega_B h^2$ and $t_0$ are only weakly dependent on 
the data set combination considered. The data considered here favors
lower $\Omega_B h^2 \sim 0.005-0.009$, or younger, $t_0 \sim 12-15$ Gyr,
universes.

Some of the constraints derived here are mildly inconsistent with those derived
elsewhere, but gratifyingly not so at high statistical significance.
Tighter and more robust constraints on cosmological parameters will require 
a models-based joint analysis of a larger collection of CMB anisotropy data
sets. 

\bigskip

We acknowledge valuable assistance from R. Stompor. This work was 
partially carried out at the California Institute of Technology
IPAC and JPL, under a contract with NASA. PM, BR, and TS acknowledge 
support from NSF CAREER grant AST-9875031. NS acknowledges support from the
Alexander von Humboldt Foundation and Japanese Grant-in-Aid for 
Science Research Fund No. 14540290.


\begin{figure}[p]
\psfig{file=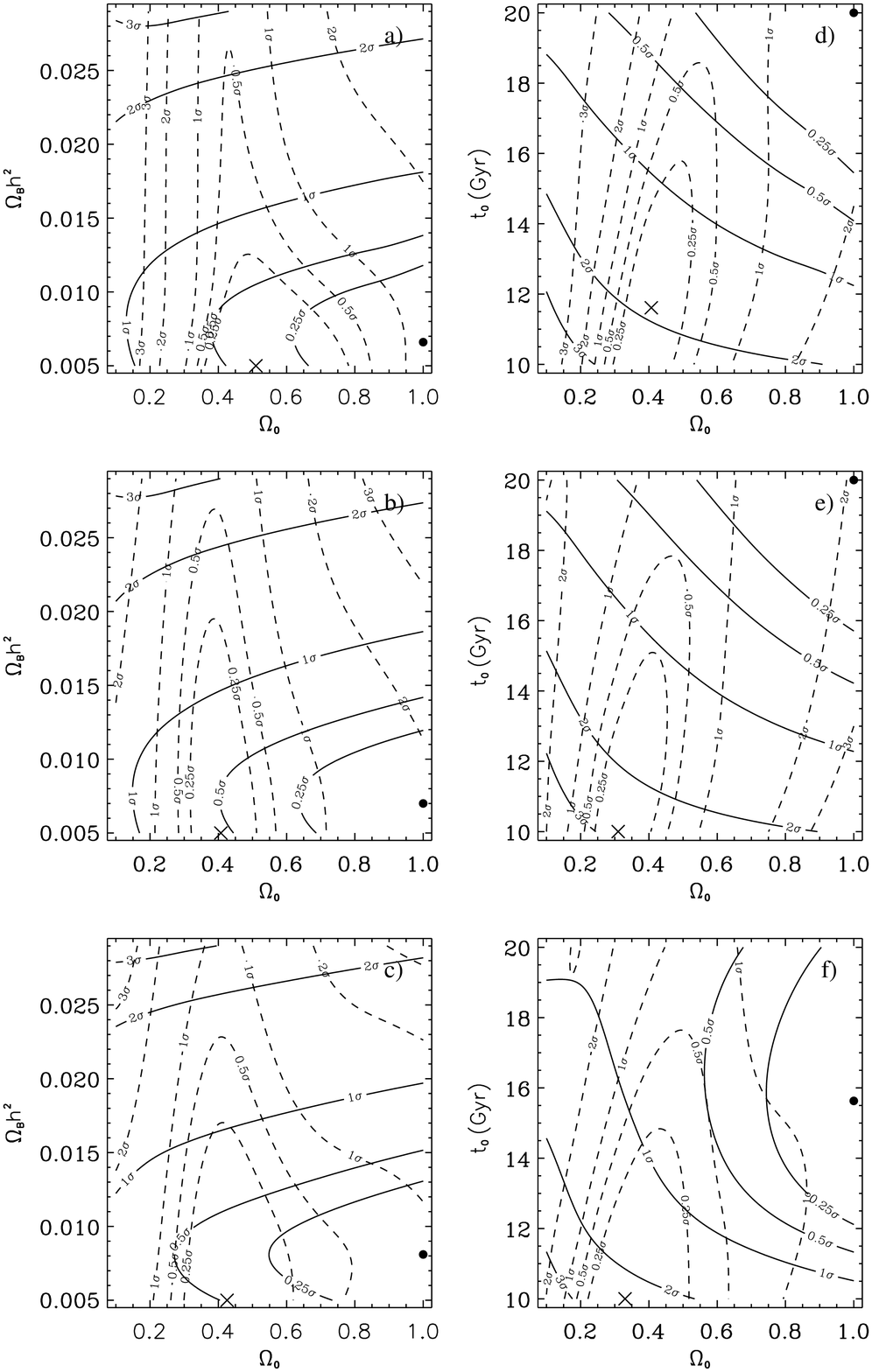,height=6.7in,angle=0}
\caption{Confidence contours and maxima of the two-dimensional posterior 
probability density distribution functions, as a function of the two 
parameters on the axes of each panel (derived by marginalizing the 
four-dimensional posterior distributions over the other two parameters). 
Dashed lines (crosses) show the contours (maxima) of the open case and 
solid lines (solid circles) show those of the flat-$\Lambda$ model. 
Contours of 0.25, 0.5, 1, 2, and 3 $\sigma$ confidence are shown
(some contours are not labelled). Panels $a)-c)$ in the left 
column show the $(\Omega_B h^2,\ \Omega_0)$ plane, while panels 
$d)-f)$ in the right column show the $(t_0, \ \Omega_0)$ plane. Panels
$a)$ \&\ $d)$ in the top row are from an analysis of all the small 
scale data. Panels $b)$ \&\ $e)$ in the middle row are from an analysis of   
all but the SuZIE small scale data. Panels $c)$ \&\ $f)$ in the bottom row 
are from an analysis of the SP94Ka, Python I--III, MAX 4 ID, MAX 5 HR,
and OVRO data sets.}
\end{figure}

\begin{figure}[p]
\psfig{file=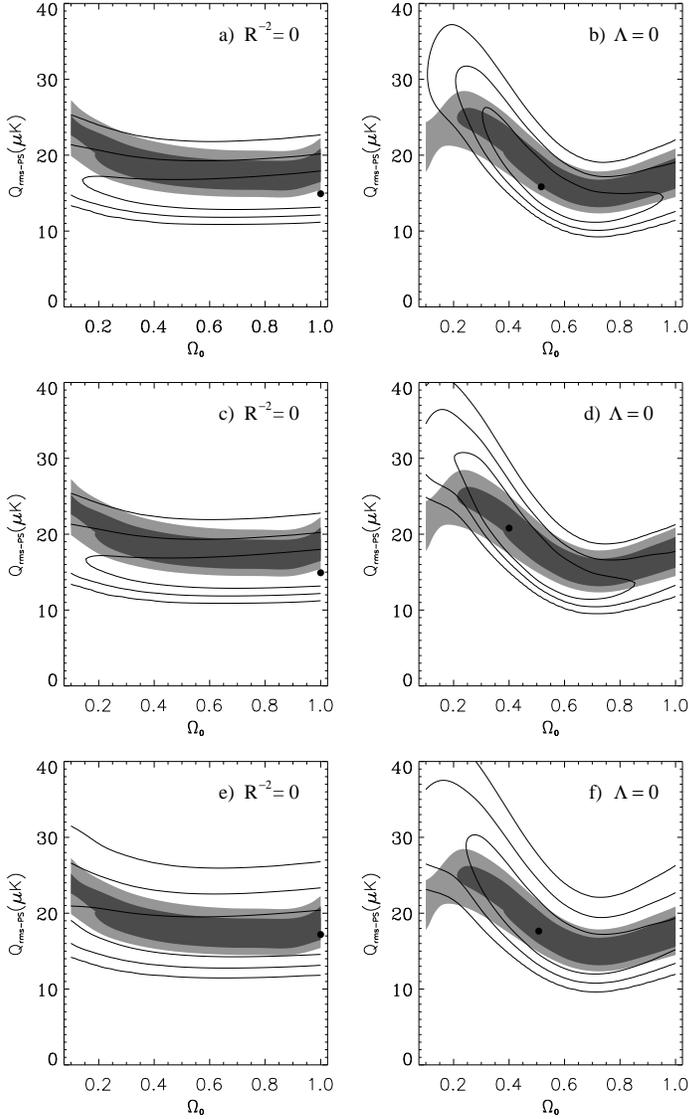,height=6.1in,angle=0}
\caption{Confidence contours and maxima of the marginalized two-dimensional 
$(Q_{\rm rms-PS}, \Omega_0)$ posterior probability density distribution 
functions. Panels $a)$, $c)$, \&\ $e)$ in the left column show the 
flat-$\Lambda$ model and panels $b)$, $d)$, \&\ $f)$ in the right column 
show the open model. Lines show the two-dimensional posterior probability 
density distribution function 1, 2, and 3 $\sigma$ confidence limits  
derived from the various combination smaller-angular-scale data sets 
considered. Solid circles show the maxima of the two-dimensional posterior 
distributions. Shaded regions show the two-dimensional posterior 
probability density distribution function confidence limits for the 
DMR data (G98; Stompor 1997); densest shading shows the 1 $\sigma$ 
confidence region and less-dense shading shows the 2 $\sigma$ region.
The DMR results are a composite of those from analyses of the two extreme 
data sets: i) galactic frame with quadrupole included and correcting for 
faint high-latitude galactic emission; and ii) ecliptic frame with 
quadrupole excluded and no other galactic emission correction (G98).
Panels $a)$ \&\ $b)$ in the top row are from an analysis of all the small 
scale data. Panels $c)$ \&\ $d)$ in the middle row are from an analysis of   
all but the SuZIE small scale data. Panels $e)$ \&\ $f)$ in the bottom row 
are from an analysis of the SP94Ka, Python I-III, MAX 4 ID, MAX 5 HR, and OVRO
data sets.}
\end{figure}

\begin{figure}[p]
\psfig{file=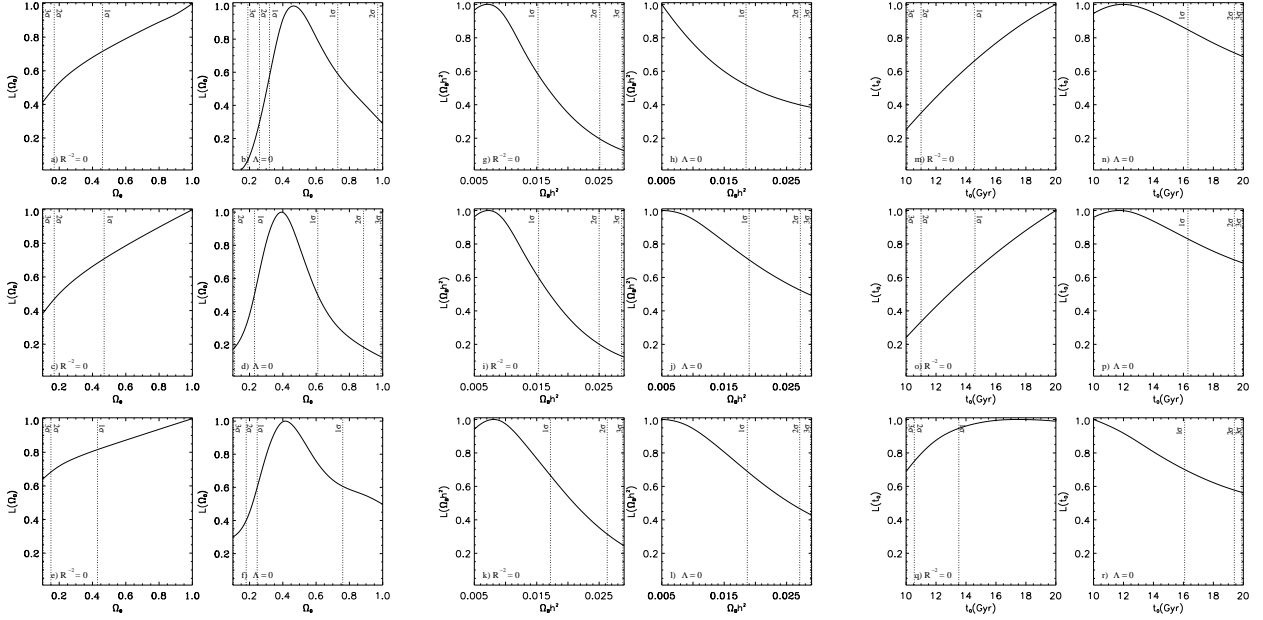,width=6.7in,angle=0}
\caption{One-dimensional posterior probability density distribution 
functions for $\Omega_0$ (two left most columns), for $\Omega_B h^2$, (two
central columns), and for $t_0$ (two right most columns), derived by 
marginalizing the four-dimensional ones over the other three parameters, 
in the open and flat-$\Lambda$ models. These have been renormalized to 
unity at the peaks. Dotted lines show the formal 1, 2, and 3 $\sigma$ 
confidence limits derived from these one-dimensional posterior distributions.
Panels $a)$, $c)$, \&\ $e)$ in the first column, $g)$, $i)$, \&\ $k)$ in
the third column, and $m)$, $o)$, \&\ $q)$ in the fifth column show the 
flat-$\Lambda$ model and panels $b)$, $d)$, \&\ $f)$ in the second column,
$h)$, $j)$, \&\ $l)$ in the fourth column, and $n)$, $p)$, and $r)$ in the
sixth column show the open model. Panels in the top row are from an 
analysis of all the small scale data, those in the middle row are from an 
analysis of all but the SuZIE small scale data, and those in the bottom 
row are from an analysis of the SP94Ka, Python I-III,
MAX 4 ID, MAX 5 HR, and OVRO data sets. When the four-dimensional posterior 
distributions are normalized such that $L(Q_{\rm rms-PS}\ =\ 0\ \mu{\rm K})\ 
=\ 1$, the peak values of the one-dimensional distributions shown in panels 
$a)-f)$ are $6\times 10^{271}$, $2\times 10^{272}$, $3\times 10^{271}$, 
$2\times 10^{272}$, $1\times 10^{202}$, and $2\times 10^{202}$, respectively;
in panels $g)-l)$ are $3\times 10^{273}$, $7\times 10^{273}$, $1\times 
10^{273}$, $6\times 10^{273}$, $5\times 10^{203}$, and $7\times 10^{203}$, 
respectively; and in  panels $m)-r)$ are $6\times 10^{270}$, $1\times 10^{271}$,
$3\times 10^{270}$, $1\times 10^{271}$, $9\times 10^{200}$, and $1\times 
10^{201}$, respectively.}
\end{figure}

\begin{figure}[p]
\psfig{file=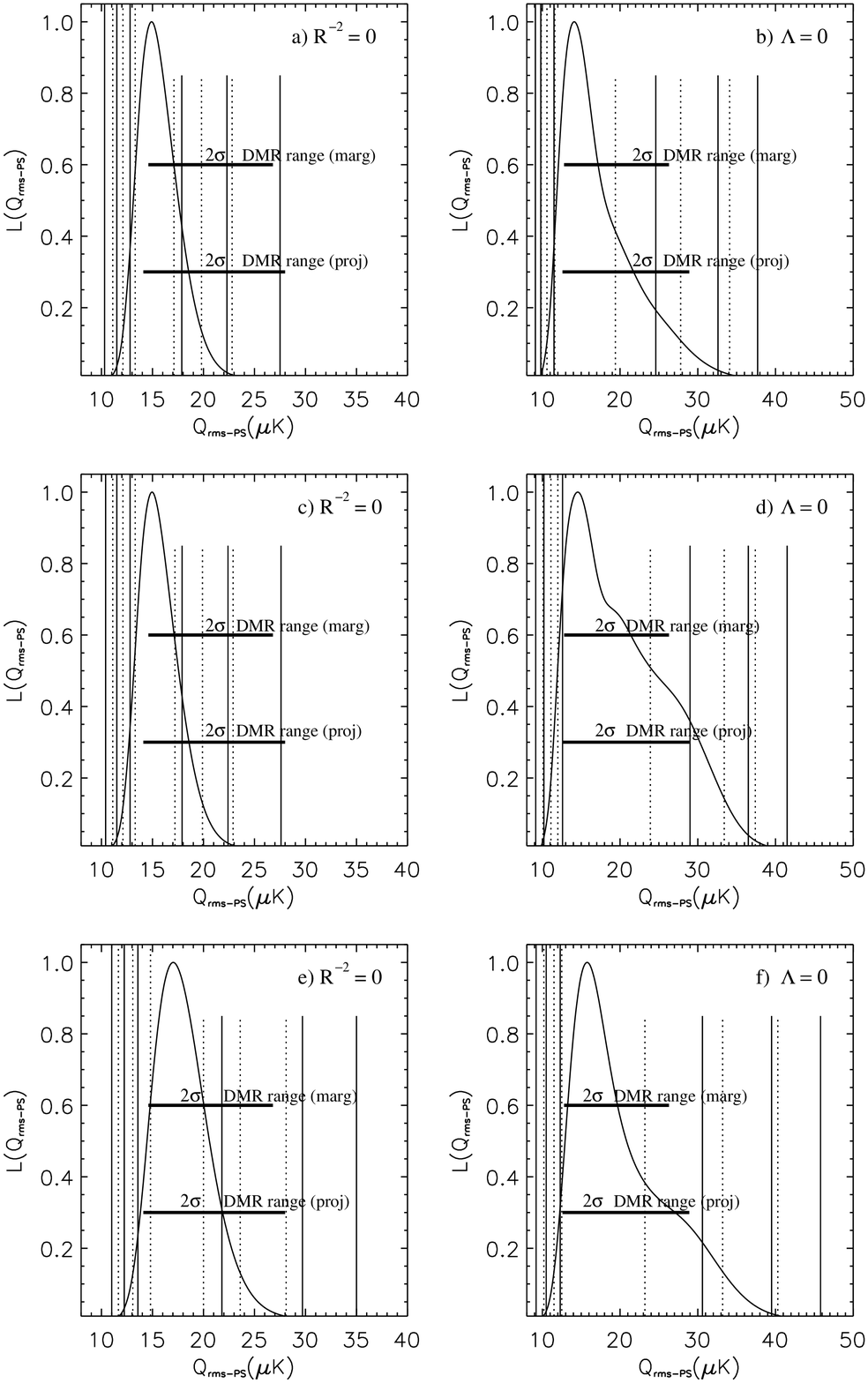,height=6.7in,angle=0}
\caption{One-dimensional posterior probability density distribution functions 
for $Q_{\rm rms-PS}$, derived by marginalizing the four-dimensional 
ones over the other three parameters, in the open and flat-$\Lambda$ models.
Conventions are as described in the caption of Fig.~3. Solid vertical lines 
show the $\pm 1$, $\pm 2$, and $\pm 3$ $\sigma$ confidence limits derived by 
projecting the corresponding small-scale data four-dimensional posterior 
distributions. Also shown are the 2 $\sigma$ DMR (marginalized and projected) 
confidence limits; these are a composite of those from the two extreme DMR 
data sets (see caption of Fig.~2). When the four-dimensional posterior 
distributions are normalized such that $L(Q_{\rm rms-PS}\ =\ 0\ \mu{\rm K})\ 
=\ 1$, the peak values of the one-dimensional distributions shown in panels 
$a)-f)$ are $9\times 10^{270}$, $1\times 10^{271}$, $4\times 10^{270}$, 
$8\times 10^{270}$, $1\times 10^{201}$, and $1\times 10^{201}$, respectively.}
\end{figure}

\begin{figure}[p]
\psfig{file=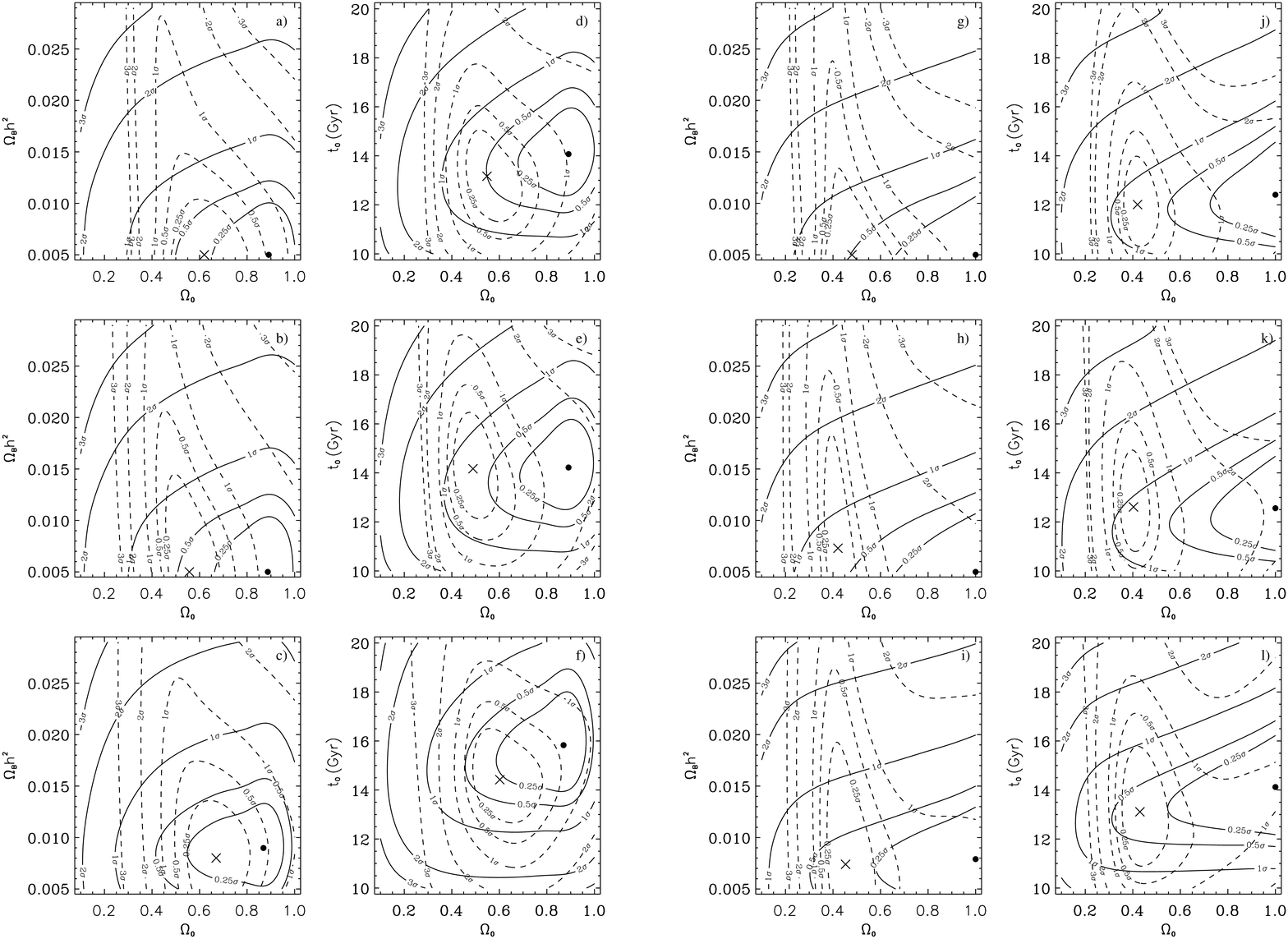,width=6.7in,angle=0}
\caption{Confidence contours and maxima of the two-dimensional posterior 
probability density distribution functions, as a function of the two 
parameters on the axes of each panel (derived by marginalizing the 
four-dimensional posterior distributions over the other two parameters). 
Dashed lines (crosses) show the contours (maxima) of the open case and 
solid lines (solid circles) show those of the flat-$\Lambda$ model. 
Contours of 0.25, 0.5, 1, 2, and 3 $\sigma$ confidence are shown
(some contours are not labelled). Panels $a)-c)$ in the first 
column and panels $g)-i)$ in the third column show the $(\Omega_B h^2,\ 
\Omega_0)$ plane, while panels $d)-f)$ in the second column and panels 
$j)-l)$ in the fourth show the $(t_0, \ \Omega_0)$ plane. Panels $a)-f)$
in the two left hand columns use the DMR galactic frame data with 
quadrupole included and corrected for faint high-latitude galactic 
emission while panels $g)-l)$ in the two right hand columns use the
DMR ecliptic frame data with quadrupole excluded and no other galactic 
emission correction (G98). Panels in the top row are from an analysis 
of the DMR and all smaller-angular-scale data, those in the middle row 
are from an analysis of all but the SuZIE data, and those in 
the bottom row are from an analysis of the DMR, SP94Ka, Python I--III, 
MAX 4 ID, MAX 5 HR, and OVRO data sets.}
\end{figure}

\begin{figure}[p]
\psfig{file=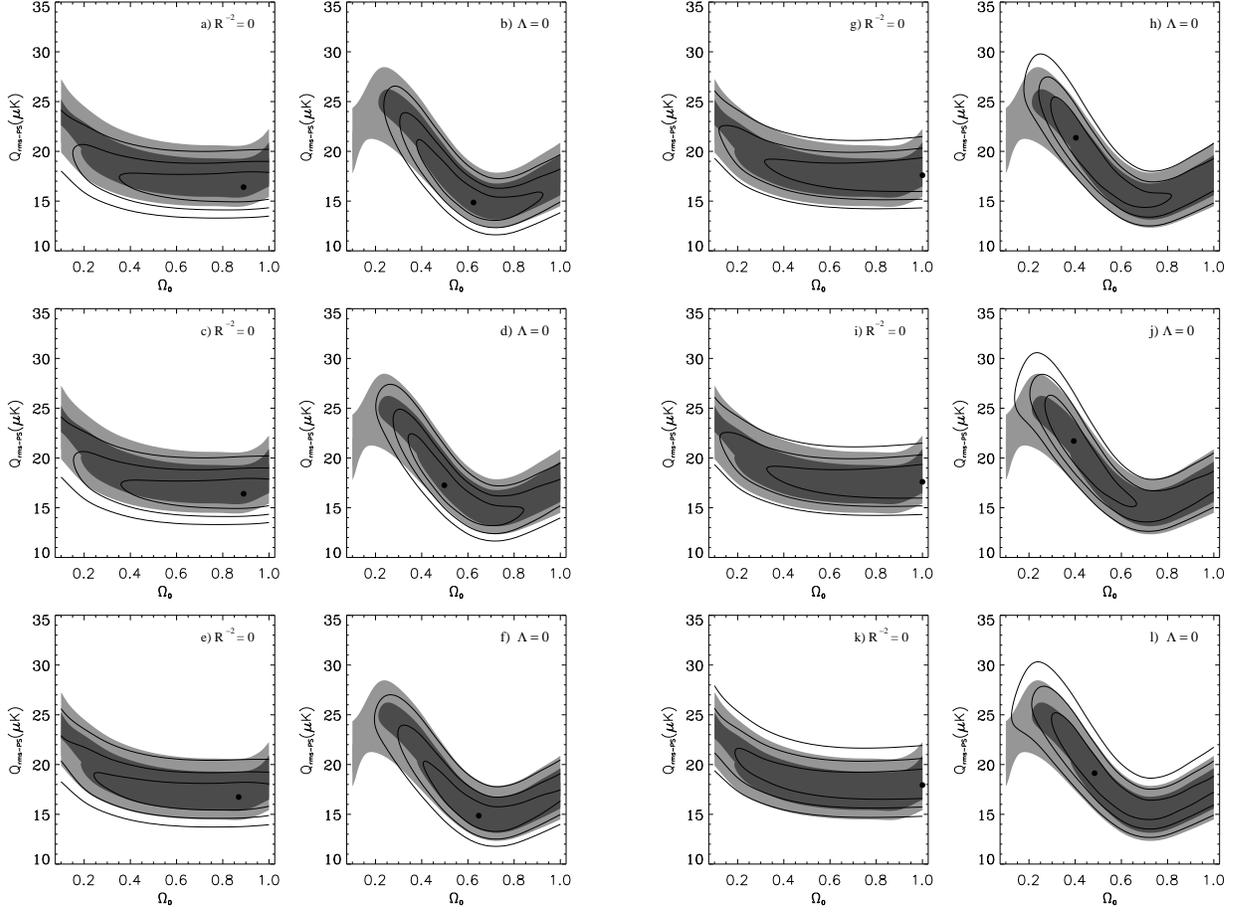,width=6.7in,angle=0}
\caption{Confidence contours and maxima of the marginalized two-dimensional 
$(Q_{\rm rms-PS}, \Omega_0)$ posterior probability density distribution 
functions. Panels $a)$, $c)$, \&\ $e)$ in the first column and
$g)$, $i)$, \&\ $k)$ in the third column show the flat-$\Lambda$ model 
while panels $b)$, $d)$, \&\ $f)$ in the second column and $h)$, $j)$,
\&\ $l)$ in the fourth column show the open model. Shaded regions show the 
two-dimensional posterior probability density distribution function 
confidence limits for the DMR data (see Fig.~2 caption for details).
Lines show the two-dimensional posterior probability density distribution 
function 1, 2, and 3 $\sigma$ confidence limits derived from the various combination data sets considered. Solid circles show the maxima of the 
two-dimensional posterior distributions. Lines in panels $a)-f)$ in the 
two left hand columns use the DMR galactic frame data with quadrupole 
included and corrected for faint high-latitude galactic emission while 
lines in panels $g)-l)$ in the two right hand columns use the DMR 
ecliptic frame data with quadrupole excluded and no other galactic 
emission correction (G98). Panels in the top row are from an analysis 
of the DMR and all smaller-angular-scale data, those in the middle row 
are from an analysis of all but the SuZIE data, and those in 
the bottom row are from an analysis of the DMR, SP94Ka, Python I--III, 
MAX 4 ID, MAX 5 HR, and OVRO data sets.}
\end{figure}

\begin{figure}[p]
\psfig{file=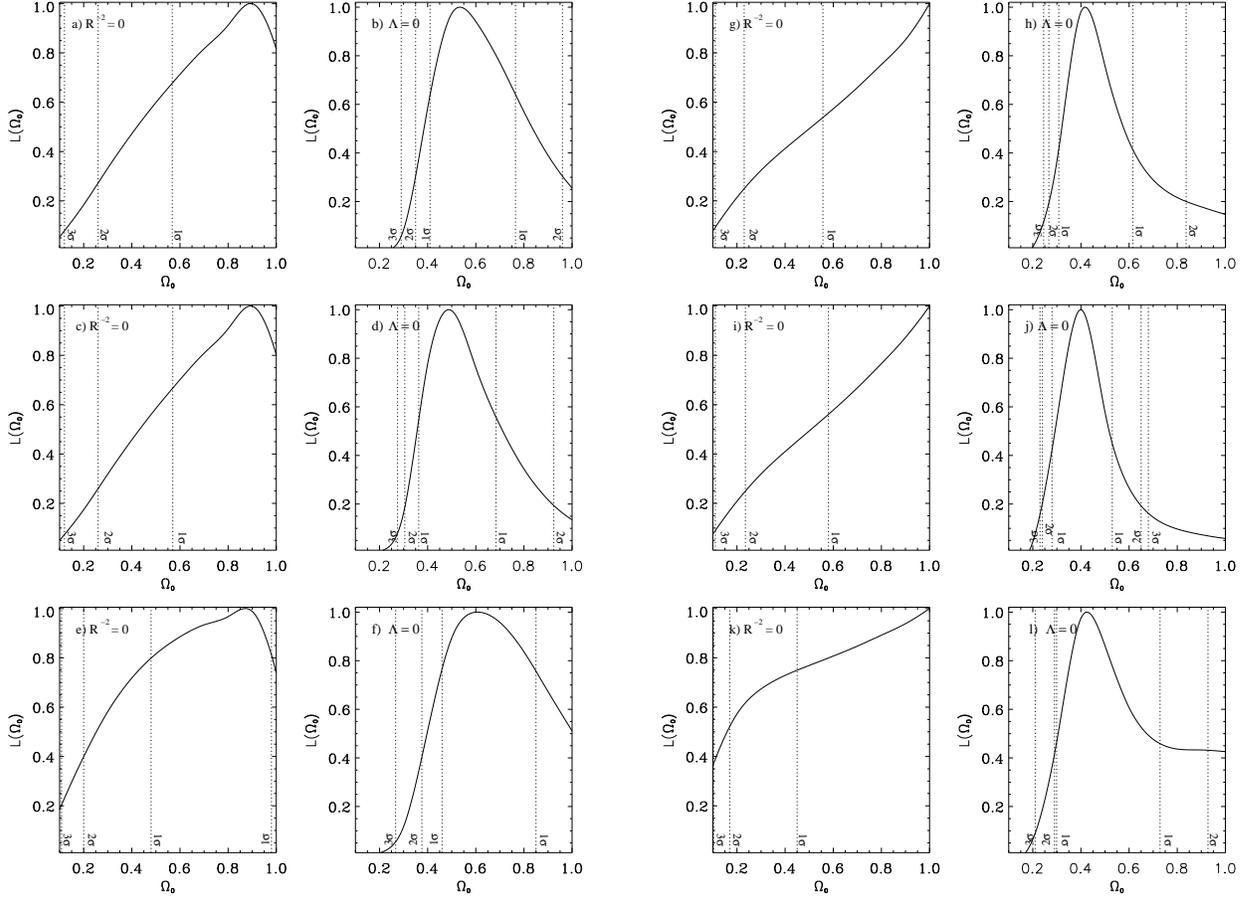,width=6.7in,angle=0}
\caption{One-dimensional posterior probability density distribution 
functions for $\Omega_0$, derived by marginalizing the four-dimensional ones 
over the other three parameters, in the open and flat-$\Lambda$ models. These 
have been renormalized to unity at the peaks. Dotted lines show the formal 1, 
2, and 3 $\sigma$ confidence limits derived from these one-dimensional
posterior distributions. Panels $a)$, $c)$, \&\ $e)$ in the first column, 
and $g)$, $i)$, \&\ $k)$ in the third column show the flat-$\Lambda$ model 
while panels $b)$, $d)$, \&\ $f)$ in the second column, and $h)$, $j)$, 
\&\ $l)$ in the fourth column show the open model. Panels $a)-f)$ in the 
two left hand columns use the DMR galactic frame data with quadrupole 
included and corrected for faint high-latitude galactic emission while 
lines in panels $g)-l)$ in the two right hand columns use the DMR 
ecliptic frame data with quadrupole excluded and no other galactic 
emission correction (G98). Panels in the top row are from an analysis 
of the DMR and all smaller-angular-scale data, those in the middle row 
are from an analysis of all but the SuZIE data, and those in 
the bottom row are from an analysis of the DMR, SP94Ka, Python I--III, 
MAX 4 ID, MAX 5 HR, and OVRO data sets. When the four-dimensional posterior 
distributions are normalized such that $L(Q_{\rm rms-PS}\ =\ 0\ \mu{\rm K})\ 
=\ 1$, the peak values of the one-dimensional distributions shown in panels 
$a)-l)$ are $5\times 10^{435}$, $1\times 10^{436}$, $2\times 10^{435}$, 
$1\times 10^{436}$, $1\times 10^{366}$, $1\times 10^{366}$, $6\times 10^{440}$, 
$4\times 10^{441}$, $3\times 10^{440}$, $6\times 10^{441}$, $2\times 10^{371}$, 
and $4\times 10^{371}$, respectively.}
\end{figure}

\begin{figure}[p]
\psfig{file=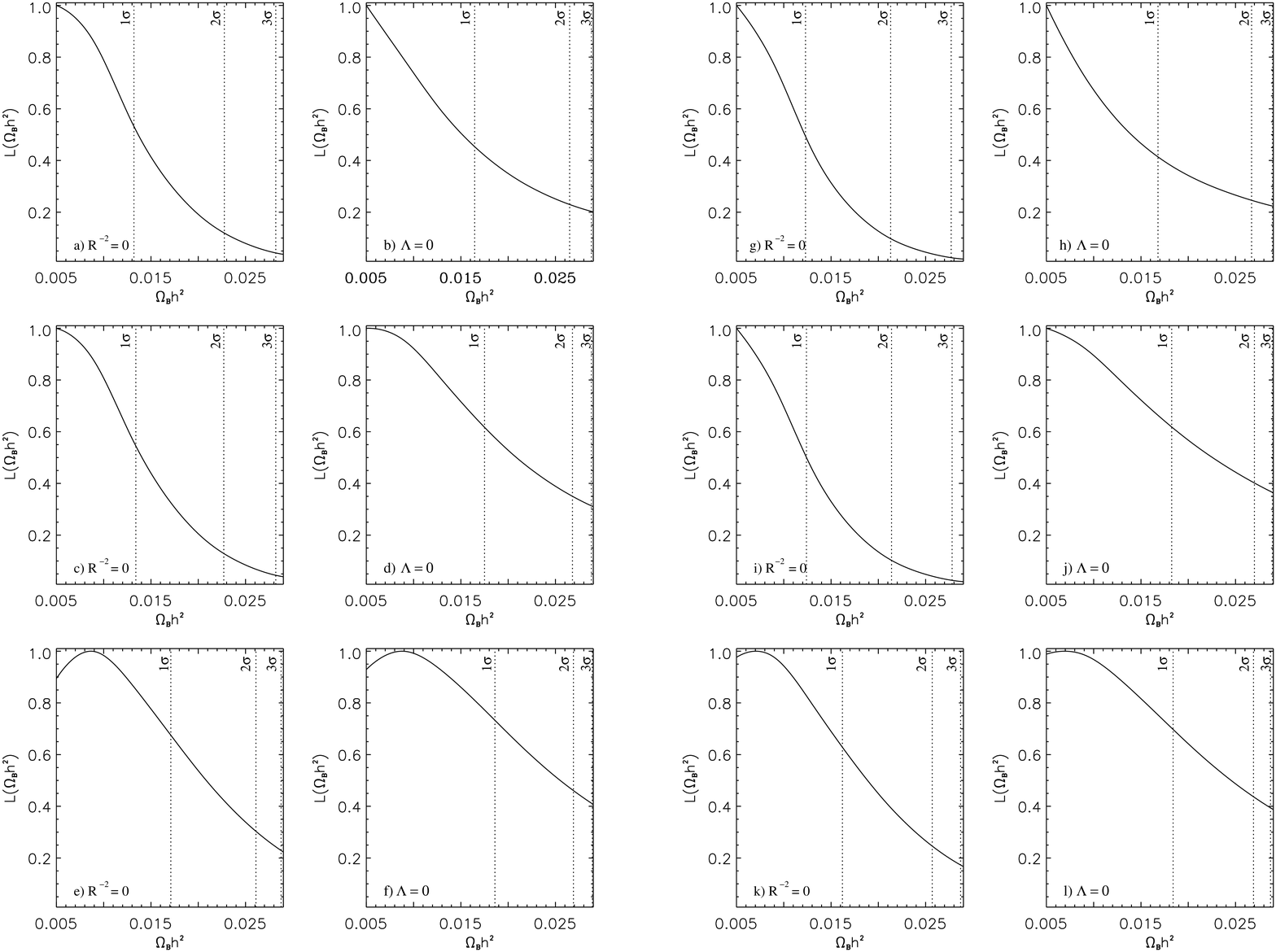,width=6.7in,angle=0}
\caption{One-dimensional posterior probability density distribution 
functions for $\Omega_B h^2$, derived by marginalizing the four-dimensional 
ones over the other three parameters, in the open and flat-$\Lambda$ models. 
Conventions are as described in the caption of Fig.~7. When the 
four-dimensional posterior distributions are normalized such that 
$L(Q_{\rm rms-PS}\ =\ 0\ \mu{\rm K})\ =\ 1$, the peak values of the 
one-dimensional distributions shown in panels $a)-l)$ are $2\times 10^{437}$, 
$6\times 10^{437}$, $1\times 10^{437}$, $3\times 10^{437}$, $4\times 10^{367}$, 
$4\times 10^{367}$, $3\times 10^{442}$, $1\times 10^{443}$, $2\times 10^{442}$, 
$9\times 10^{442}$, $8\times 10^{372}$, and $1\times 10^{373}$, respectively.}
\end{figure}

\begin{figure}[p]
\psfig{file=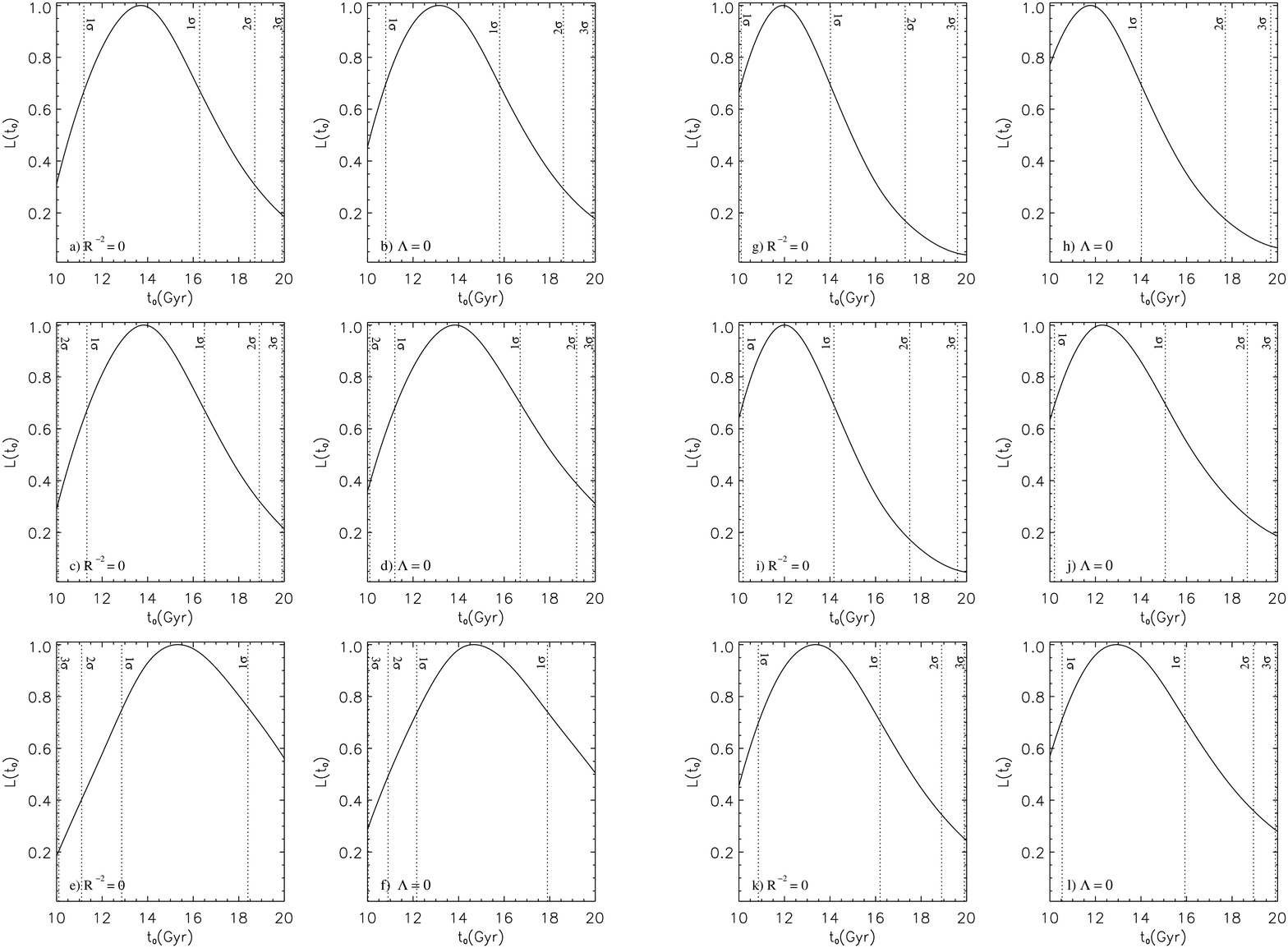,width=6.7in,angle=0}
\caption{One-dimensional posterior probability density distribution 
functions for $t_0$, derived by marginalizing the four-dimensional 
ones over the other three parameters, in the open and flat-$\Lambda$ models. 
Conventions are as described in the caption of Fig.~7. When the 
four-dimensional posterior distributions are normalized such that 
$L(Q_{\rm rms-PS}\ =\ 0\ \mu{\rm K})\ =\ 1$, the peak values of the 
one-dimensional distributions shown in panels $a)-l)$ are $4\times 10^{434}$, 
$1\times 10^{435}$, $2\times 10^{434}$, $8\times 10^{434}$, $9\times 10^{364}$, 
$9\times 10^{364}$, $6\times 10^{439}$, $3\times 10^{440}$, $3\times 10^{439}$, 
$2\times 10^{440}$, $2\times 10^{370}$, and $3\times 10^{370}$, respectively.}
\end{figure}

\begin{figure}[p]
\psfig{file=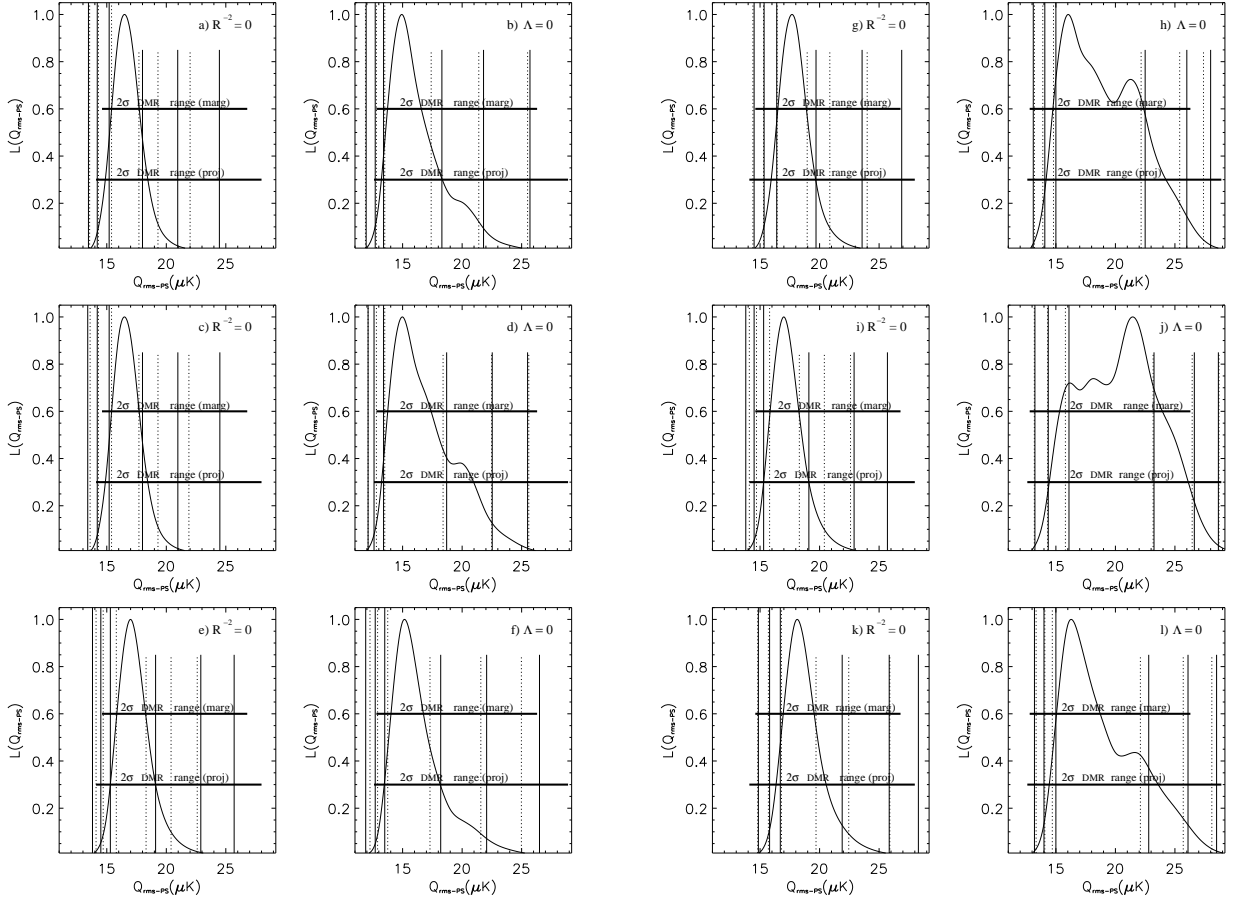,width=6.7in,angle=0}
\caption{One-dimensional posterior probability density distribution 
functions for $Q_{\rm rms-PS}$, derived by marginalizing the four-dimensional 
ones over the other three parameters, in the open and flat-$\Lambda$ models. 
Conventions are as described in the caption of Fig.~7. Also shown here as 
solid vertical lines are $\pm 1$, $\pm 2$, and $\pm 3$ $\sigma$ confidence 
limits derived by projecting the corresponding four-dimensional posterior 
distributions, as well as the 2 $\sigma$ DMR (marginalized and projected) 
confidence limits which are a composite of those from the two extreme DMR 
data sets (see caption of Fig.~2). When the 
four-dimensional posterior distributions are normalized such that 
$L(Q_{\rm rms-PS}\ =\ 0\ \mu{\rm K})\ =\ 1$, the peak values of the 
one-dimensional distributions shown in panels $a)-l)$ are $8\times 10^{434}$, 
$1\times 10^{435}$, $4\times 10^{434}$, $1\times 10^{435}$, $2\times 10^{365}$, 
$2\times 10^{365}$, $1\times 10^{440}$, $2\times 10^{440}$, $5\times 10^{439}$, 
$2\times 10^{440}$, $4\times 10^{370}$, and $3\times 10^{370}$, respectively.}
\end{figure}

\end{document}